\documentclass[aps,twocolumn,a4paper,showkeys,showpacs,floatfix]{revtex4-1}
\usepackage{graphicx}
\usepackage{amsmath,amsfonts}
\usepackage{subfigure}
\begin{document}
\title{A Quantum Model for Coherent Ising Machines: \\ Stochastic Differential Equations with Replicator Dynamics}
\author{Taime Shoji}
\author{Kazuyuki Aihara}
\affiliation{Institute of Industrial Science, the University of Tokyo, 4-6-1 Komaba, Meguro-ku, Tokyo, 153-8505, Japan}
\author{Yoshihisa Yamamoto}
\affiliation{E.L.Ginzton Laboratory, Stanford University,Stanford,CA94305, USA}
\begin{abstract}
The quantum theory of coherent Ising machines, based on degenerate optical parametric oscillators and measurement-feedback circuits, is developed using the positive $P(\alpha,\beta)$ representation of the density operator and the master equation. The theory is composed of the c-number stochastic differential equations  for describing open dissipative quantum dynamics and the replicator dynamics equations for handling measurement-induced collapse of the density operator. We apply the present theory to simulate two simple Ising spin models and elucidate the unique features of this computing machine.
\end{abstract}
\maketitle
\section{Introduction}
Various combinatorial optimization problems belong to the NP-hard or NP-complete class and are difficult to solve in a polynomial time with a deterministic Turing machine\cite{NPhard}. To find exact or approximate but satisfying solutions for such problems, many heuristic algorithms, such as classical neural networks\cite{Hopfield}\cite{HopfieldTank}, simulated annealing\cite{SimulatedAnnealing}, and quantum annealing\cite{Farhi}\cite{QuantumAnnealing}, are proposed.

The Ising problem is the simplest model for finding the minimum energy ground state of spin glasses\cite{IsingOriginal}. The Ising Hamiltonian is given by
\begin{equation}
\mathcal H = \sum_{i,j \in E} J(i,j)\sigma(i)\sigma(j),
\end{equation}
where $G = (V,E)$ is a given graph, $J: E\to \mathbb R$ is the weight of an edge, and $\sigma:V\to \{-1,1\}$ is the vertex value, called an ``Ising spin.'' It is known that a three-dimensional Ising model and two-dimensional Ising model with a Zeeman field are NP-hard problems.\cite{Ising}

There have been several attempts to solve Ising problems with actual physical devices, called Ising machines, rather than algorithms inspired by physical phenomena. A coherent Ising machine (CIM) is one such physical device\cite{InjectionCIM1}\cite{InjectionCIM2}\cite{OPOCIM1}. The first generation of CIM uses injection-locked lasers to represent Ising spins,\cite{InjectionCIM1}\cite{InjectionCIM2} while the second generation employs degenerate optical parametric oscillators (DOPOs)\cite{OPOCIM1}. The coupling $J(i,j)$ between Ising spins is implemented by the optical delay line coupling between the oscillators\cite{NaturePhotonics}\cite{TakataSciRep}\cite{InagakiNature}.

Recently, a measurement-feedback circuit has been used to implement $J(i,j)$ efficiently. The exprimental system with N DOPO pulses shown in Fig.1(a) \cite{Science1}\cite{Science2} is modeled and sinplified as shown in Fig 1(b). In this study, we develop the quantum theory of such measurement-feedback based CIMs based on the positive $P(\alpha,\beta)$ representation of the density operator and the c-number stochastic differential equations (CSDE) with replicator dynamics.  The accompanying paper addresses the same problem with a different theoretical model based on a completely positive trace preserving map\cite{Yamamura}.

\section{Master equation for CIM}
In this section, we will derive the master equation of a measurement-feedback-based CIM. For the sake of simplicity, we describe the derivation of the master equation for only a two-spin system, but the theory can be easily extended to many-spin systems.
\subsection{A simple model of CIM}
\begin{figure}[htbp]
        \subfigure{
        \includegraphics[width=90mm,clip]{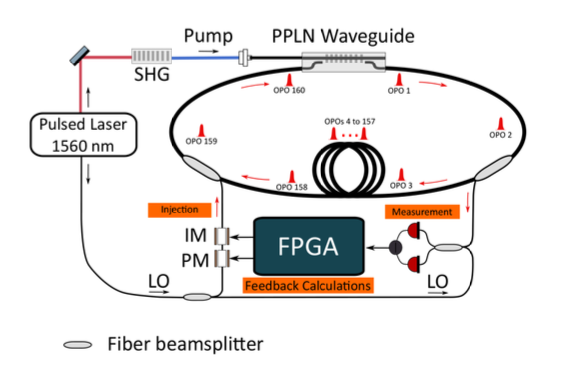}
        }
        \subfigure{
        \includegraphics[width=90mm,clip]{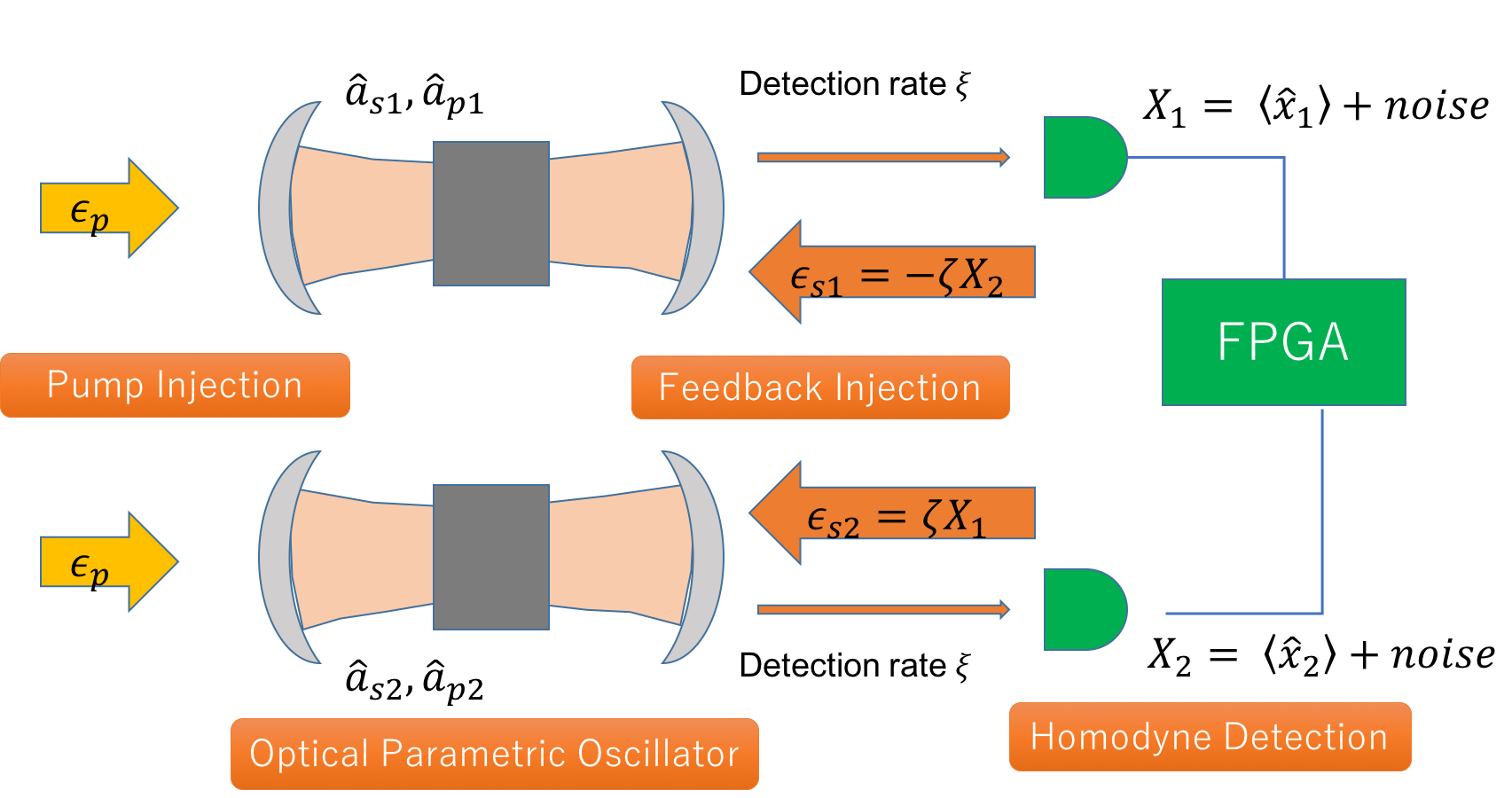}
        }
        \caption{(Color online) (a)The experimental setup of CIM\cite{Science1}\cite{Science2}. (b) A Simplified model o f CIM.} \label{fig:CIM}
\end{figure}

Figure 1(b) shows a simplified model of the CIM. There are two DOPO cavities with the identical signal frequency $\omega_s$ and pump frequency $\omega_p = 2\omega_s$. The photon annihilation operators of the signal and pump fields are denoted by $\hat{a}_{s1},\hat{a}_{p1},\hat{a}_{s2}$, and $\hat{a}_{p2}$. There are also two external fields injected into the cavities. One is an excitation pump field at $\omega_p$. Another is a feedback signal field at $\omega_s$, which is prepared by the measurement-feedback circuit.

The intra-cavity pump field and signal field have the loss rates denoted by $\gamma_p$ and $\gamma_s$, respectively. To measure the in-phase component $\hat X_i = \hat a_{si} + \hat a_{si} ^\dag (i = 1,2) $ of the signal field, a part of the intra-cavity field is picked off and measured by homodyne detectors. The feedback signal is prepared based on the measurement results.

Since the measurement-feedback process is local operation and classical communication (LOCC), the density matrix of the total system stays in a product state during the whole computation process and is given as follows:
\begin{eqnarray}
\rho = \rho_1 \otimes \rho_2,
\end{eqnarray}
where $\rho_i$ is a density matrix of the DOPO $i$.

Our theretical model is a continuous time evolution model, in which all the quantum operations proceed simultaneously, while a real measurement-feedback-based CIM \cite{Science1}\cite{Science2} is based on discrete quantum operations, as shown in Fig. 1(a). Extension of the present work to describe a discrete model is straightforward and will be discussed elsewhere.

In a real CIM, each DOPO takes the form of a pulse circulating in a fiber ring resonator and is operated sequentially in time. Let $\tau$ be time interval of DOPO pulses and consider a set of DOPO pulse amplitudes $\{\hat a_{s1}(t), \hat a_{s2}(t+\tau), ... , \hat a_{sN}(t+(N-1)\tau)\}$, which appear at a same spatial point (injection coupler) in a real experimental system. The internal target pulse and externally prepared feedback pulse collide at the injection coupler with the exactly same delay time, so that a delay of feedback signal in the theoretical morel can be assumed to be zero.

\subsection{Derivation of the master equation}
To acquire the stochastic differential equations for the CIM, we treat a DOPO and a measurement-feedback circuit separately.

The Hamiltonian of DOPOs is given by 
\begin{eqnarray}
\hat H &=&\hat H _{free} + \hat H _{int} + \hat H _{pump}  + \hat H _{FB}+ \hat H_{loss},\\
\hat H_{free} &=& \hbar\sum_{i=1,2} \omega_{s}\hat a^\dag_{si} \hat a _{si} +  \omega_{p}\hat a^\dag_{pi} \hat a_{pi},\\
\hat H_{int} &=& \frac{i\hbar\kappa}{2} \sum_{i=1,2}\hat a_{si}^{\dag 2 }\hat a_{pi} -\hat a_{si}^2 \hat a_{pi}^\dag ,,\\
\hat H_{pump} &=& i\hbar \sum_{i=1,2} \epsilon_p \hat a_{pi}^\dag e^{-i\omega_d t} - \epsilon_p^* \hat a_{pi} e^{i\omega_d t},\\
\hat H_{FB} &=& i\hbar \sum_{i=1,2} \epsilon_s \hat a_{pi}^\dag e^{-i\omega_f t} - \epsilon_s^* \hat a_{pi} e^{i\omega_f t},\\
\hat H_{loss} &=& \hbar \sum_{i=1,2} \hat a_{si} ^\dag \hat \Gamma_{si} + \hat a_{si} \hat \Gamma_{si}^\dag + a_{pi} ^\dag \hat \Gamma_{pi} + a_{pi} \hat \Gamma_{pi}^\dag,
\end{eqnarray}
where $\kappa$ is a parametric coupling constant between the signal field and the pump field in a nonlinear crystal, and $\Gamma_{pi}$ and $\Gamma_{si}$ are the external reservoir field operators, which account for the fluctuation forces injected from the external environment. By tracing out these external fields by the standard Born--Markov approximation\cite{Gerdiner}, we can obtain the following master equation of the DOPOs:

\begin{eqnarray}
\frac{d \hat \rho_{DOPO}}{dt} &=& \sum_{i=1,2}i\hbar\omega_s[\hat a_{si}^\dag \hat a_{si}, \hat \rho] + i\hbar\omega_p[\hat a_{pi}^\dag \hat a_{pi},\hat \rho]\nonumber\\
&+& \frac{\gamma_s}{2}(2\hat a_{si} \hat \rho \hat  a_{si}^\dag - \hat a_{si}^\dag \hat a _{si}\hat \rho - \hat \rho \hat a_{si}^\dag \hat a_{si})\nonumber  \\
&+&\frac{\gamma_p}{2}(2\hat a_{pi} \hat \rho \hat a_{pi}^\dag - \hat a_{pi}^\dag \hat a _{pi}\hat \rho - \hat \rho \hat a_{pi}^\dag \hat a_{pi})\\
&+& [\epsilon_{si}e^{-i\omega t} \hat a_{si}^\dag - \epsilon_{si}^* e^{i\omega t} \hat a_{si}^\dag , \hat \rho] \nonumber\\
&+& [\epsilon_{pi}e^{-i\omega t} \hat a_{si}^\dag - \epsilon_{pi}^* e^{i\omega t} \hat a_{pi}^\dag , \hat \rho] \nonumber\\
&+& \frac{i\hbar\kappa}{2}[\hat a^{\dag2}_{si}\hat a_{pi} - \hat a_{si}^2 \hat a_{pi}^\dag, \hat \rho].\nonumber
\end{eqnarray}
By taking a rotating reference frame properly, we can eliminate the two terms in the first line of Eq.(9).

To describe the nonunitary reduction of a wave function by the homodyne measurement of $\hat x$, Wiseman and Milburn proposed the following equation\cite{WisemanFeedback}:
\begin{eqnarray}
\frac{d \hat \rho_{meas}}{dt} &=& \sum_{i=1,2}\frac{\xi}{2}(2\hat a_{si} \hat \rho \hat  a_{si}^\dag - \hat a_{si}^\dag \hat a _{si}\hat \rho - \hat \rho \hat a_{si}^\dag \hat a_{si})  \nonumber\\
&+&\sqrt{\xi} \frac{dW}{dt} \left(\hat a_{si}\rho+\hat\rho \hat a_{si}^\dag - \left< \hat a_{si} + \hat a_{si}^\dag\right>\hat\rho\right),
\end{eqnarray}
where $\left<\hat X_i\right> = \left<\hat a_{si}^\dag + \hat a_{si}\right>$ is the expectation value of the in-phase amplitude of the signal field and $dW$ is the Wiener increment, which satisfies
\begin{eqnarray}
dW_i(t) &\sim& \mathcal N(0, dt),\\
\left< dW_i(t) dW_j(t') \right> &=& 2\pi \delta_{ij}\delta(t - t').
\end{eqnarray}
 In this model, the actually measured value $X_i$ is given by
\begin{eqnarray}
X_i dt = \left<\hat a_{si}^\dag + \hat a_{si}\right> dt + \frac{dW_i}{\sqrt\xi}.
\end{eqnarray}
A feedback signal $\epsilon_{si}$ is now prepared according to the formula,
\begin{eqnarray}
\epsilon_{si} = \zeta \sum_{j} J_{ij} X_j,
\end{eqnarray}
where $\zeta$ is the strength of the feedback coupling between DOPOs.

Finally, we obtain the overall master equation of the measurement-feedback-based CIM by combining Eq.(9) and (10) as follows:

\begin{eqnarray}
\frac{d \hat \rho}{d t} &=& \frac{\gamma_s+\xi}{2}(2\hat a_{si} \hat \rho \hat  a_{si}^\dag - \hat a_{si}^\dag \hat a _{si}\hat \rho - \hat \rho \hat a_{si}^\dag \hat a_{si})\nonumber \\
&+& \frac{\gamma_p}{2}(2\hat a_{pi} \hat \rho \hat a_{pi}^\dag - \hat a_{pi}^\dag \hat a _{pi}\hat \rho - \hat \rho \hat a_{pi}^\dag \hat a_{pi})\nonumber\\
&+& [\epsilon_{si}e^{-i\omega t} \hat a^\dag - \epsilon_{si}^* e^{i\omega t} \hat a_{si}^\dag , \hat \rho] \nonumber\\
&+& [\epsilon_{pi}e^{-i\omega t} \hat a^\dag - \epsilon_{pi}^* e^{i\omega t} \hat a_{pi}^\dag , \hat \rho] \\
&+& \frac{i\hbar\kappa}{2}[\hat a^{\dag2}_{si}\hat a_{pi} - \hat a_{si}^2 \hat a_{pi}^\dag, \hat \rho]\nonumber \\
&+& \sqrt{\xi} \frac{dW}{dt} \left(\hat a_{si}^\dag\hat \rho +\hat \rho\hat a_{si} - \left< \hat a_{si}^\dag + \hat a_{si}\right>\hat\rho\right).\nonumber
\end{eqnarray}

\section{Stochastic differential equations}
\subsection{Positive $P(\alpha,\beta)$ representation}
 The $P(\alpha,\beta)$ representation of the density operator is defined by\cite{OPOhamiltonian}
  \begin{equation}
    \hat{\rho} = \int\int P(\alpha,\beta) \hat{\Lambda}(\alpha,\beta)d^2\alpha d^2\beta,
  \end{equation}
  where $\alpha,\beta \in \mathbb C$ and
  \begin{equation}
    \hat{\Lambda}(\alpha,\beta) = \frac{|\alpha\left.\right >\left<\right.\beta |}{\left<\alpha |\beta \right>}
  \end{equation}
  is the off-diagonal projector in terms of coherent states. Here, $ |{\bf \alpha} \left.\right>$ and $|{\bf \beta} \left. \right>$ are the tensor product coherent states: $|{\bf \alpha} \left.\right>= |\alpha_{s1}\left.\right>\otimes|\alpha_{p1} \left.\right>\otimes|\alpha_{s2} \left.\right>\otimes|\alpha_{p2} \left.\right>$, and $|{\bf \beta} \left.\right> = |\beta_{s1} \left.\right>\otimes|\beta_{p1} \left.\right>\otimes|\beta_{s2} \left.\right>\otimes|\beta_{p2} \left.\right>$.

An important property of the positive $P(\alpha,\beta)$ representation is that we can always define $P({\bf \alpha}, {\bf \beta})$ as  a positive and real function for arbitrary quantum states of fields, which satisfies
\begin{eqnarray}
\int d^2\alpha d^2\beta P(\alpha,\beta) = 1.
\end{eqnarray}
 Therefore, we can regard $P(\alpha,\beta)$ as a probability distribution function for finding the projector $|\alpha\left.\right>\left<\right.\beta |$ in the density matrix. 
 
 Since a coherent state is an eigenstate of an annihilation operator, the moment of the density matrix is easily evaluated by
 \begin{eqnarray}
 \left< \hat a^{\dag n} \hat a^m \right> = \int d^2\alpha d^2\beta  \alpha^m\beta^nP(\alpha,\beta).
 \end{eqnarray}
 Using this relation, we can obtain any statistics of an observable, which is composed by the creation and annihilation operators such as $\left<\hat X\right> = \left<\hat a + \hat a^\dag\right>$.
 
 The probability density function of finding the in-phase component $X$ can be also computed from the positive P representation as follows:
 \begin{eqnarray}
 Prob(X) &=&Tr [\left. |X\right>\left< X|\right. \hat \rho] \\\nonumber
 &=& \sqrt{\frac{2}{\pi}}\int d^2\alpha d^2\beta \exp\left[2\left(X-\frac{\alpha+\beta}{2}\right)\right]P(\alpha,\beta).
 \end{eqnarray}

 Because the density matrix of the measurement-feedback-based CIM is given as the product state of each DOPO density matrix, we can express $P(\alpha,\beta)$ as
 \begin{eqnarray}
 P(\alpha,\beta) &=& P(\alpha_{s1},\beta_{s1},\alpha_{p1},\beta_{p1})\nonumber\\
 &&\times P(\alpha_{s2},\beta_{s2},\alpha_{p2},\beta_{p2}).
 \end{eqnarray}
 Therefore, we can describe the total system with the partial differential equation (PDE) for each DOPO.

Using the properties of the coherent states, we obtain the PDE of $P(\alpha_{si},\beta_{si},\alpha_{pi},\beta_{pi})$ from the master equation as follows:
\begin{eqnarray}
\frac{\partial P(\alpha,\beta)}{\partial t} &=& \left[\sqrt\xi \left\{ \alpha_{si} + \beta_{si} - \left<\alpha_{si} + \beta_{si}\right> \right\}\frac{dW}{dt} \right.\nonumber \\
&-& \frac{\partial}{\partial \alpha_{si}} \left\{ -(\gamma_s+\xi) \alpha_{si} + \kappa\beta_{si}\alpha_{pi} + \epsilon_{si}\right\}\nonumber \\
&-&\frac{\partial}{\partial\beta_{si}}\left\{-(\gamma_s+\xi) \beta_{si} + \kappa\alpha_{si}\beta_{pi} + \epsilon_{si}\right\}\nonumber\\
&-&\frac{\partial}{\partial \alpha_{pi}}\left\{-\gamma_p\alpha_{pi} -\frac{\kappa}{2}\alpha_{si}^2+\epsilon_{pi}\right\} \\
&-& \frac{\partial}{\partial\beta_{pi}}\left \{-\gamma_p\beta_{pi} - \frac{\kappa}{2}\beta_{si}^2+\epsilon_{pi}\right\}\nonumber\\
&+&\left. \frac{\partial^2}{\partial \alpha_{si}^2} \kappa\alpha_{pi} + \frac{\partial^2}{\partial\beta_{si}^2}\kappa\beta_{pi} \right]P(\alpha,\beta),\nonumber
\end{eqnarray}
where
\begin{eqnarray} 
\left< \alpha_{si} \right> = \int d^2\alpha d^2\beta \alpha_{si} P(\alpha,\beta),\\ \left< \beta_{si} \right> =\int d^2 \alpha d^2\beta \beta_{si}P(\alpha,\beta).
\end{eqnarray}

\subsection{Stochastic differential equations and replicator dynamics}
Except for the first line in Eq. (22), the PDE has an identical form as the Fokker--Planck equation. It is well established that the Fokker--Planck equation can be transformed to the following stochastic differential equations (SDE)\cite{Gerdiner}:
\begin{eqnarray}
\left[ \begin{array}{c} d\alpha_{si} \\ d\beta_{si}\end{array} \right]
&=& \left[ \begin{array}{c}
-(\gamma_s+\xi)\alpha_{si} + \kappa \beta_{si} \alpha_{pi} + \epsilon_{si} \\
-(\gamma_s+\xi)\beta_{si} + \kappa\alpha_{si} \beta_{pi} + \epsilon_{si}
\end{array} \right]dt\nonumber\\&+&
\left[ \begin{array}{cc}
\kappa \alpha_{pi} & 0 \\ 0 &\kappa\beta_{pi}
\end{array}\right]^{1/2}
\left[\begin{array}{c} dW_{\alpha_i} \\ dW_{\beta_i} \end{array}\right],\\
\left[ \begin{array}{c}d \alpha_{pi} \\ d\beta_{pi}\end{array} \right]
&=& \left[ \begin{array}{c}
-\gamma_p\alpha_{pi} + \frac{\kappa}{2}\alpha_{si}^2 + \epsilon_{pi} \\
-\gamma_p\beta_{pi} + \frac{\kappa}{2} \beta_{si}^2 + \epsilon_{pi}
\end{array} \right]dt.
\end{eqnarray}
When $\gamma_p >> \gamma_s$, the pump field decays more rapidly than the signal field, so that the pump field follows the dynamics of the signal field (the slaving principle). We can eliminate the pump field by assuming $d\alpha_{pi} = d\beta_{pi} = 0$,
\begin{eqnarray}
\left[ \begin{array}{c} d\alpha_{si} \\ d\beta_{si}\end{array} \right]
&=& \left[ \begin{array}{c}
-(\gamma_s+\xi)\alpha_{si} + \frac{\kappa}{\gamma_p}\beta_{si} (\epsilon_{pi} - \frac{\kappa}{2}\alpha_{si}^2) + \epsilon_{si} \\
-(\gamma_s+\xi)\beta_{si} + \frac{\kappa}{\gamma_p}\alpha_{si} (\epsilon_{pi} - \frac{\kappa}{2}\beta_{si}^2) + \epsilon_{si}
\end{array} \right]dt \nonumber\\&+&
\left[ \begin{array}{c}
\sqrt{\frac{\kappa}{\gamma_p}(\epsilon_{pi} - \frac{\kappa}{2}\alpha_{si}^2)}dW_{\alpha_i}\\
\sqrt{\frac{\kappa}{\gamma_p}(\epsilon_{pi} - \frac{\kappa}{2}\beta_{si}^2)}dW_{\beta_i}
\end{array}\right].
\end{eqnarray}
By introducing $\eta_i = g\alpha_{si}, \mu_i = g\beta_{si}, d\tau = \gamma_s dt, d\omega_{\eta_i} = \sqrt\gamma_s dW_{\alpha_i}, d\omega_{\mu_i} = \sqrt\gamma_s dW_{\beta_i}, p_i = \kappa\epsilon_{pi}/\gamma_s\gamma_p, f_i = g\epsilon_{si}/\gamma_s$, and $g = \kappa/\sqrt{2\gamma_p\gamma_s}, \xi' = \xi/\gamma_s$, we obtain the normalized SDE
\begin{eqnarray}
\left[ \begin{array}{c} d\eta_{i} \\ d\mu_{i}\end{array} \right]
= \left[ \begin{array}{c}
-(1+\xi')\eta_{i} + \mu_{i} (p_i - \eta_{i}^2) + f_i \\
-(1+\xi')\mu_{i} + \eta_i(p_i - \mu_{i}^2) + f_i
\end{array} \right]d\tau \nonumber \\
 +
\left[ \begin{array}{c}
g\sqrt{p_i - \eta_i^2}d\omega_{\eta_i}\\
g\sqrt{p_i -\mu_i^2}d\omega_{\mu_i}
\end{array}\right].
\end{eqnarray}
Note that above the oscillation threshold, fixed points of in-phase component $\left< \hat X_i \right> = \left<\alpha_i + \beta_i\right>$ are $\pm \sqrt{p-1}/g$, and that of mean photon number $\left<\hat n \right>$ is $(p-1)/g^2$. Therefore, to tune the present theoritical model to a real experiment, we can set $g$ from the measured photon number at a pump rate $p=2$.

On the other hand, the first line of Eq. (22), which describes the reduction of the signal density operator induced by the measurement, cannot be simulated by the standard method using the SDE. In previous works\cite{WisemanFeedback}, by assuming that the measurement result is incidentally identical to the expectation value, they ignored this term. In this study, we need to know the measurement effect on the evolution of the DOPO state, so that we keep the random measurement results on $\eta_i(t)$ and $\mu_i(t)$ by taking pseudo-random numbers $dW$.

 Because the first line of Eq. (22) is a replicator equation, we extend the branching Brownian motion model \cite{BranchingBrownianMotion}, which is called replicator dynamics in our case. In replicator dynamics, the change of $P(\alpha,\beta)$ is governed by
  \begin{eqnarray}
  \frac{\partial P(\alpha,\beta)}{\partial t} = \lambda(\alpha,\beta) P(\alpha,\beta).
  \end{eqnarray}
  Here a Brownian particle at $(\alpha,\beta)$ is
 \begin{eqnarray}
 \left\{ \begin{array}{ll}
 	\mathrm{copied\ with\ probability\ } \lambda(\alpha,\beta), & (\lambda(\alpha,\beta) > 0),\\
	\mathrm{vanished\ with\ probability\ } -\lambda(\alpha,\beta), & (\lambda(\alpha,\beta) < 0) ,
	\end{array}\right.\nonumber \\
 \end{eqnarray}
 where $\lambda(\alpha,\beta) = X_i - \left<\hat X_i\right>$.

 Because the expectation values $\left< \alpha_{si} + \beta_{si} \right>$ are needed to compute $\lambda(\alpha,\beta)$, we run many Brownian particles according to the identical SDEs and the same measurement results $X_i$ at the same time.

\subsection{Gaussian approximation}
In this section, we derive an approximation method to describe the measurement-feedback-based DOPO system.

We start from the following PDE of the signal fields after the adiabatic elimination of the pump field:

\begin{eqnarray}
\frac{\partial P(\alpha,\beta)}{\partial t} &=& \left[\sqrt\xi \left\{ \alpha_{si} + \beta_{si} - \left<\alpha_{si} + \beta_{si}\right> \right\} \frac{dW}{dt}\right. \nonumber\\
&-&\frac{\partial}{\partial \alpha_{si}} \left\{ -\gamma_s \alpha_{si} + \frac{\kappa}{\gamma_p}\beta_{si}\left(\epsilon_{pi}-\frac{\kappa\alpha_{si}^2}{2}\right) +\epsilon_{si}\right\} \nonumber\\
&-&\frac{\partial}{\partial\beta_{si}}\left\{-\gamma_s \beta_{si} + \frac{\kappa}{\gamma_p}\alpha_{si}\left(\epsilon_{pi}-\frac{\kappa\beta_{si}^2}{2}\right) + \epsilon_{si}\right\}\nonumber\\
&+& \frac{\partial^2}{\partial \alpha_{si}^2}\frac{\kappa}{\gamma_p}\beta_{si}\left(\epsilon_{pi}-\frac{\kappa\alpha_{si}^2}{2}\right) \\
&+& \frac{\partial^2}{\partial\beta_{si}^2}\left.\frac{\kappa}{\gamma_p}\alpha_{si}\left(\epsilon_{pi}-\frac{\kappa\beta_{si}^2}{2}\right) \right]P(\alpha,\beta).\nonumber
\end{eqnarray}
By partial integration of Eq. (31), the equations of motion for the expectation values are obtained as follows:

\begin{eqnarray}
 d\left< \alpha_{si} \right> & =& \sqrt\xi \left[\left< \alpha_{si}^2 \right> + \left<\alpha_{si}\beta_{si}\right> -\left<\alpha_{si}\right>^2 -\left<\alpha_{si}\right> \left<\beta_{si}\right> \right]dW\nonumber \\
&&+\left[-\gamma_s\left<\alpha_{si} \right> +\frac{\kappa\epsilon_{si}}{\gamma_p}\left<\beta_{si}\right>\right. \\
&&-\left.\frac{\kappa^2}{2\gamma_p}\left<\alpha_{si}^2\beta_{si}\right> + \epsilon_{si} \right] dt\nonumber,\\
d\left< \beta_{si} \right> &=& \sqrt\xi \left[\left< \alpha_{si}\beta_{si}\right> + \left<\alpha_{si}^2\right> -\left<\alpha_{si}\right> \left<\beta_{si}\right> -\left<\beta_{si}\right>^2 \right]dW\nonumber\\
&&+ \left[-\gamma_s\left<\beta_{si} \right> +\frac{\kappa\epsilon_{si}}{\gamma_p}\left<\alpha_{si}\right>\right. \\
&&-\left.\frac{\kappa^2}{2\gamma_p}\left<\beta_{si}^2\alpha_{si}\right> + \epsilon_{si} \right] dt.\nonumber
\end{eqnarray}
Similarly, we can derive the equations of motion for the higher order statistics such as $\left<\alpha_{si}^2\right>, \left<\beta_{si}^2\right>$, and $\left<\alpha_{si}\beta_{si}\right>$. Even though $\epsilon_{si}$ contains the statistics of other DOPOs such as $\left< \alpha_{sj} \right>(i\neq j)$, $\left< \alpha_{si} \alpha_{sj}\right>$ is expressed as $\left< \alpha_{si}\right>\left<\alpha_{sj}\right>$ since the total density matrix is separable. By changing the basis from $\left< \alpha_{si} \right> and \left< \beta_{si} \right>$ to $\left <X_i\right> = \left<\alpha_{si} \right > +\left< \beta_{si}\right> and \left< iP_i \right> = \left< \alpha_{si}\right> - \left<\beta_{si}\right>$, the equations of motion of $\left< X_i \right>$ and $\left<P_i\right>$ are acquired.

Though the dynamical equations are acquired, we cannot simulate Eq. (32) and (33) immediately because of higher order terms like $\left<\alpha_{si}^2\beta_{si}\right>$. To avoid this difficulty, we consider the approximate wave function given by a displaced squeezed vacuum state

\begin{eqnarray}
\left. |\psi_i \right> = \hat D(\mu_i)\hat S(\sigma_i^2) \left.|0\right>,
\end{eqnarray}
where $\hat D(\mu_i) = \exp(\mu_i \hat a^\dag - \mu_i^*\hat a)$ is the displacement operator, and $\hat S(\sigma_i^2) = \exp(1/2(\sigma_i^{2*} \hat a^2 - \sigma_i^2\hat a^{\dag 2}))$ is the squeezing operator. For simplicity, both $\mu$ and $\sigma_i^2$ are real. This approximation means that the DOPO state is always described by a pure squeezed state and higher order statistics has no effect on the dynamics of the system. By this approximation, we can finally get the dynamical equations of motion of the DOPOs as follows:
\begin{eqnarray}
d\mu_i &=& \sqrt\xi \left(\sigma_i^2- \frac{1}{4}\right) dW + \left[ -\gamma_s\mu + \frac{\kappa}{\gamma_p}\epsilon_p\mu \right.\\
&-&\left. \frac{\kappa^2}{2\gamma_p}\left (\mu^3 + \frac{\mu_i}{\sigma_i^2}\left(\sigma_i^2-\frac{1}{4}\right)\left(3\sigma_i^2-\frac{1}{4}\right)\right)+\epsilon_{si}\right]dt,\nonumber\\
d\sigma_i^2 &=& \left[-2\gamma_s\left(\sigma_i^2-\frac{1}{4}\right)+\frac{2\kappa}{\gamma_p}\epsilon_p\left( \sigma_i^2 + \frac{1}{4}\right)\right.\nonumber\\
 &-& \frac{\kappa^2}{2\gamma_p}\left( \frac{5}{8}+6\sigma_i^4 + 6\sigma_i^2\mu_i^2 - \frac{1}{2}\sigma_i^2 + \frac{3}{2}\mu_i^2 - \frac{3}{32\sigma_i^2}\right)\nonumber\\
  &-&\left. 4\xi\left(\sigma_i-\frac{1}{4}\right)^2\right]dt,
\end{eqnarray}
where
\begin{eqnarray}
\epsilon_{si} = \zeta \sum_j J_{ij}\left( \mu_i + \frac{dW}{\sqrt \xi}\right).
\end{eqnarray}
 The first term in Eq. (35) describes the shift of the center position of the wave function by the measurement. The last term in Eq.(36) describes the reduction of the variance by the measurement.

\section{Numerical simulation results}
In this section, we will show the numerical simulation results based on exact CSDE (28) and replicator dynamics of Eq. (29) and compare them with the simulation results by the Gaussian approximation based on Eqs. (35) and (36).
\subsection{N=2 DOPO model}

\begin{figure*}[htbp]

        \subfigure{
           \includegraphics[width=55mm,clip]{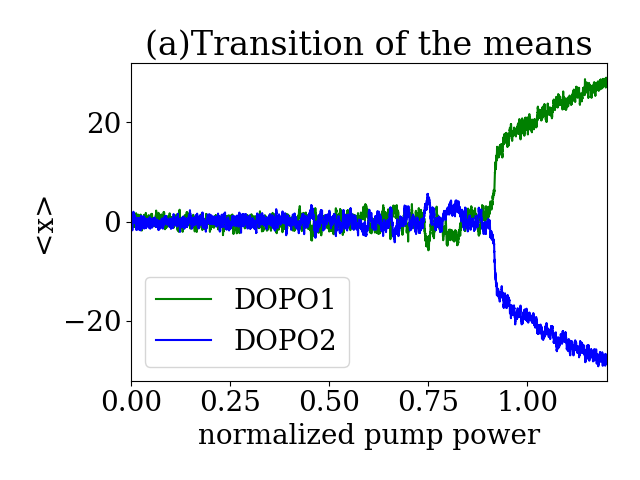}
        }
        \subfigure{
           \includegraphics[width=55mm,clip]{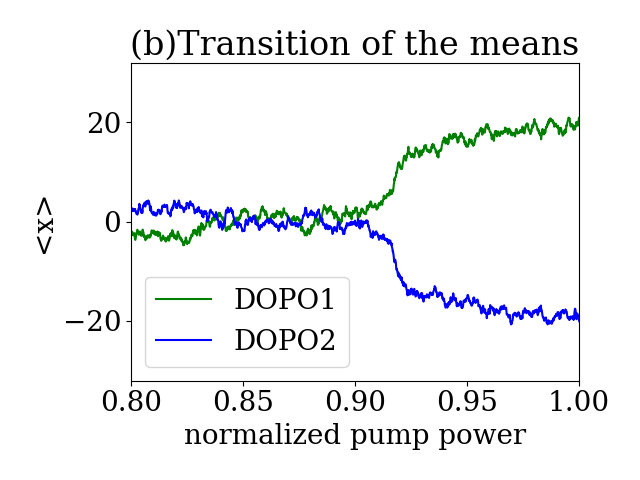}
        }
        \subfigure{
        \includegraphics[width=55mm,clip]{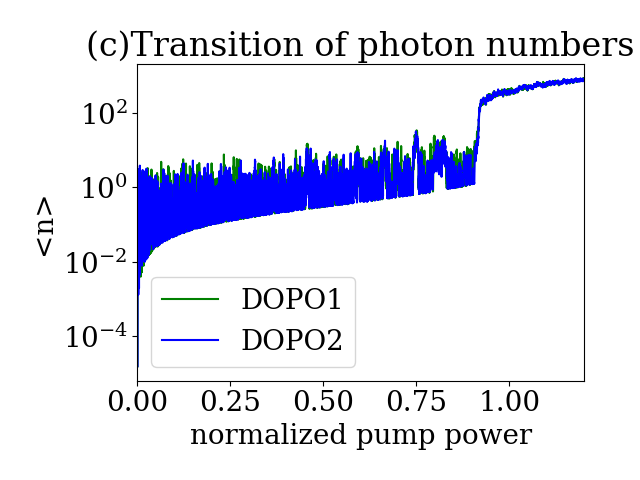}
        }
        \subfigure{
           \includegraphics[width=55mm,clip]{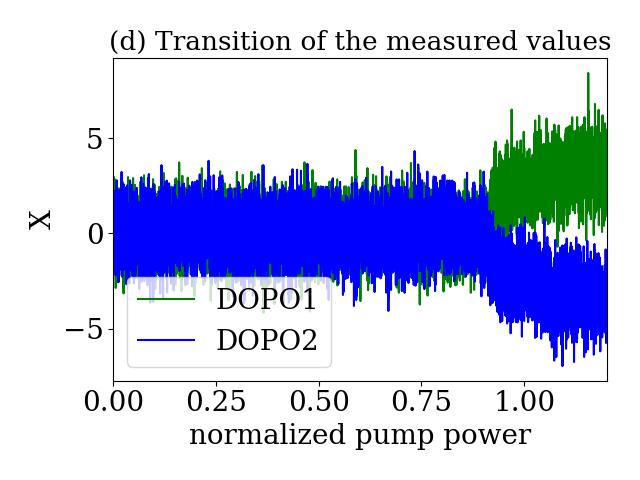}
	}
        \subfigure{
        \includegraphics[width=55mm,clip]{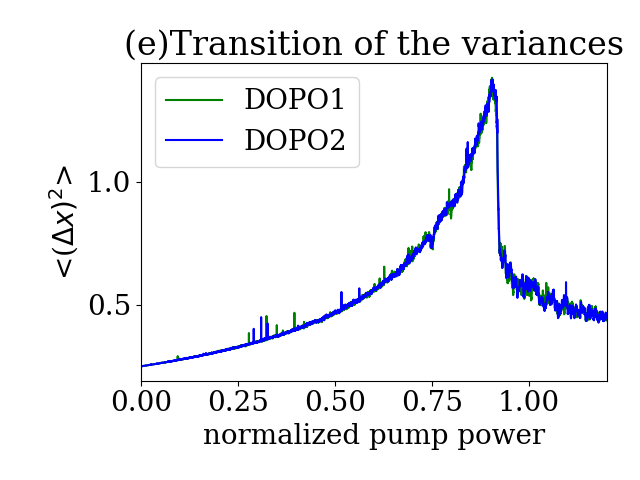}
        }
        \subfigure{
        \includegraphics[width=55mm,clip]{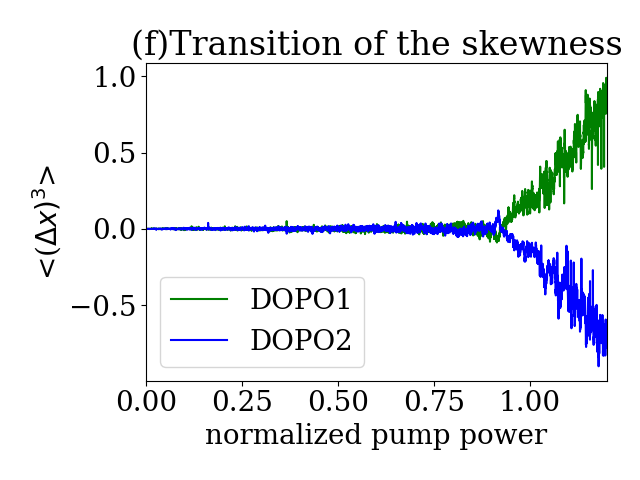}
        }
        \subfigure{
        \includegraphics[width=55mm,clip]{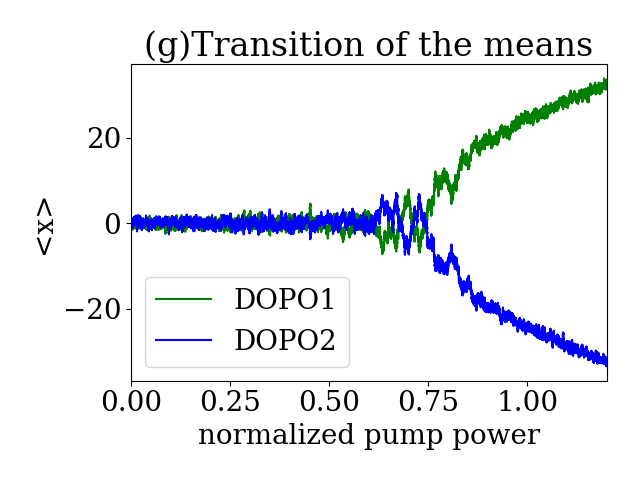}
        }
        \subfigure{
        \includegraphics[width=55mm,clip]{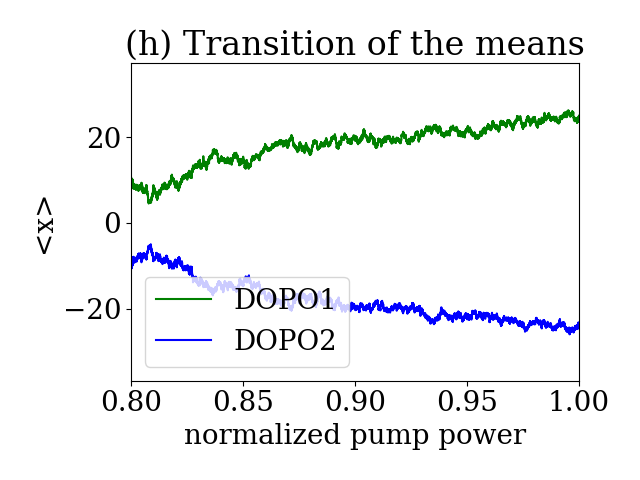}
        }
        \subfigure{
        \includegraphics[width=55mm,clip]{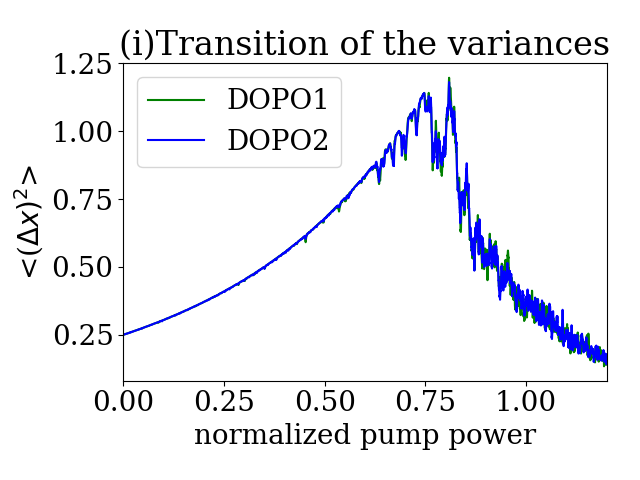}
        }
        \caption{(Color online) (a)Time evolution of the expectation values of the in-phase components $\left<X_1\right>$ (green (light gray) line) and $\left<X_2\right>$ (blue (dark gray) line). This single trajectories of $\left<X_1\right>$ and $\left< X_2 \right>$ are generated by the ensemble average over 10,000 Brownian particles. (b) Magnified picture of Fig. 2(a) near the decision making point. (c) Average photon numbers in the DOPO cavities.(d) Measured values of $X_1$ and $X_2$ by the homodyne detectors. (e) Variances $\sigma_1^2$ and $\sigma_2^2$ of the in-phase components. (f) Time evolution of the skewness of the in-phase components $\left<\Delta X^3\right>$. (g)Time evolution of the expectation values of the in-phase components $\left< X_1 \right>$ and $\left< X_2\right>$ under the Gaussian approximation. (h) Magnified view of (g) immediately above the threshold. (i) Variances of the in-phase components $\sigma_1^2$ and $\sigma_2^2$ under the Gaussian approximation.} \label{fig:sqvars}
\end{figure*}

First, we study the system of two DOPOs with antiferromagnetic coupling. Figure 2 shows the time evolution of the average in-phase amplitudes $\left < X_i \right>$ vs. normalized pump rate $\epsilon_p/\epsilon_{th}$, where $\gamma_s = 1, \gamma_p = 10, \kappa = 0.1, \xi = 0.1$, and  $\zeta = 0.3$. The corresponding saturation parameter is $g =0.02$. Figure 2(a)-(f) shows the result of exact replicator dynamics discussed in sec.II-B, and Figure 2(g)-(h) shows that of Gaussian approximation discussed in sec.II-C. The external pump rate $\epsilon_{p}$ is linearly increased from 0 to 1.5 times the threshold value $\epsilon_{th}$. The two DOPOs are coupled by the antiferromagnetic interaction ($J_{ij} = -1$). Figure 2(b) expands the average in-phase amplitudes $\left< X_1 \right>$ and $\left<X_2 \right>$ near the bifurcation point (decision making point). The two DOPOs point toward one ground state $\left. |\uparrow\downarrow\right>$ at one time but switch back to the other ground state $\left. |\downarrow\uparrow\right>$ at another time. This random search process continues until the final decision is made at $\epsilon_p/\epsilon_{th} \simeq 0.9$. 

Figure 2(c) shows the average photon number $\left<n\right>$ versus the normalized pump rate $\epsilon_p/\epsilon_{th}$. The lower bound of the average photon number below the threshold in Fig. 2(f) is associated with the squeezed vacuum state, while the noisy spikes correspond to the finite in-phase amplitude induced by the measurement-feedback process. Note that the average photon number per DOPO is on the order of one at this decision making instance, as shown in Fig.2(c).

Figure 2(d) shows the measurement results actually reported by the homodyne detectors. We conclude that the negative correlation is formed between $\left< X_1 \right>$ and $\left<X_2\right>$ at very early stages by the measurement-feedback process as shown in Fig.2(b), but the actual measurement results are too noisy to disclose those quantum search processes. A final solution, which the CIM will report should be determined at the effective threshold pump rate $ \epsilon_p/\epsilon_{th} = 1-\zeta = 0.7$. We will discuss in the next section that the late bifurcation at $\epsilon_p/\epsilon_{th} = 0.9$ in Fig.2(b) rather than $0.7$ stems from the quantum tunneling.

The variance $\left< \Delta X_i^2 \right>$ and skewness$ \left< \Delta X_i^3\right> $ in the anti-squeezed in-phase amplitudes are shown in Fig.2(e) and (f). The DOPO wavepackets near and above the threshold are clearly deviated from the Gaussian wavepackets, for which $\left<\Delta X_i^3\right>=0$ holds. This is because the bottom of the potential function, $V_b(X_i)=-\frac{1}{2}(\epsilon_p/\epsilon_{th}-1)X_i^2+\frac{1}{4}X_i^4$, above the threshold does not have a symmetric barrier, i.e. steep barrier toward a large amplitude $|X_i| \neq 0$ and gradual barrier at zero amplitude $|X_i|=0$. However, the wavepackets below the threshold differ only slightly from the Gaussian wavepackets. Figures 3(a) and 3(b) show the probabilities $P(x > 0)$ and $P(x < 0)$ of the DOPO wave functions, respectively, and compare them with those of the Gaussian wavepackets. These results suggest that the tails of the DOPO wavepackets toward $|X_i|=0$ are broader than those of the Gaussian wavepackets even at a pump rate below the threshold.

\begin{figure*}[htbp]

        \subfigure{
        \includegraphics[width=65mm,clip]{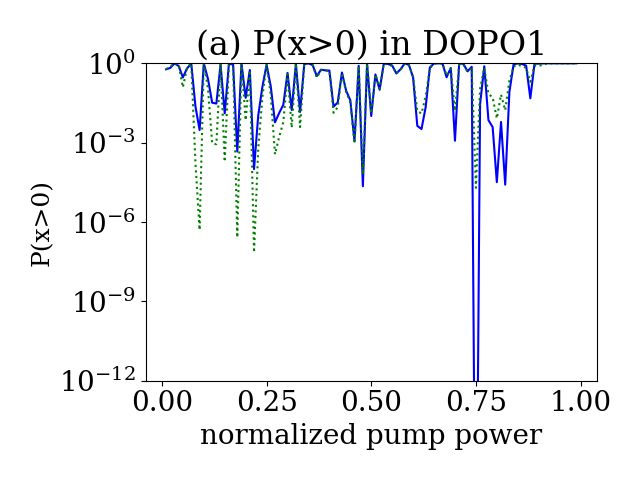}
        }
        \subfigure{
        \includegraphics[width=65mm,clip]{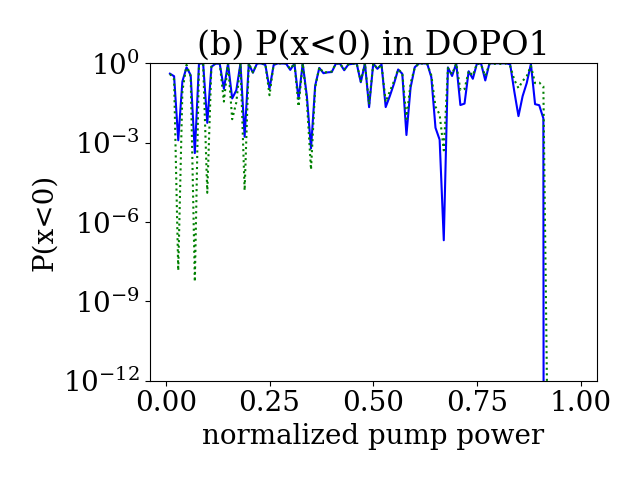}
        }
        \caption{(Color online)Time evolution of the probabilities of (a) $P(x>0)$ and (b) $P(x<0)$ in DOPO1 whose final destination is $\left.|\uparrow\downarrow\right>$ (or $\left<\hat x\right> > 0$). Blue solid lines show the probability calculated from Eq. (20), and green dotted lines show the probability calculated from the normal distribution with the same means and variances.}
\end{figure*}

 Figures 2(g) and (h) show $\left< X_i \right>$ based on the Gaussian approximation described above, and Figure 4(i) shows the variances $\sigma_i^2$s. We can find that the final correlation between $\left<X_1\right>$ and $\left<X_2\right>$ is formed already at the effective  threshold $\epsilon_p/\epsilon_{th} = 0.7$. Compared with the numerical results of the exact replicator dynamics, the bifurcation of $\left<x_1\right>$ and$ \left< x_2\right>$, namely a final decision making occurs at the threshold $\epsilon_p/\epsilon_{th}=1-\zeta = 0.7$. Please note a remarkable difference between Fig.2(b) for exact dynamics and Fig.2(h) for Gaussian approximation.

Note that the deviation from Gaussian wavepackets and the occurrence of skewness are also observed in non degenerate optical parametric oscillator (NDOPO) systems and contribute the entanglement generation in NDOPO systems\cite{Referees1}. Existence of nonlinear effect in optical systems yields non Gaussian distribution which leads to deference from classical systems\cite{Referees2}\cite{Referees3}.

\subsection{N=16 DOPO system}
 
 \begin{figure}
	\subfigure{
        \includegraphics[width=60mm,clip]{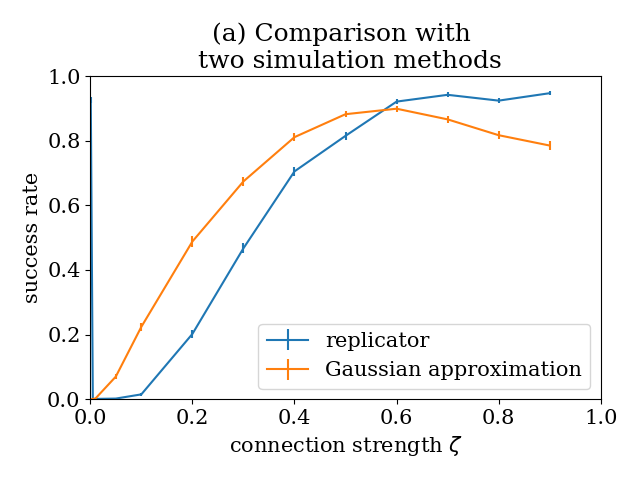}
        }
        \subfigure{
        \includegraphics[width=60mm,clip]{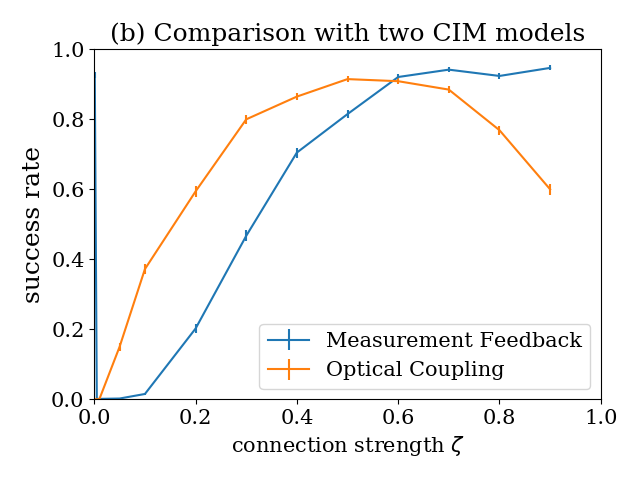}
		}
		\subfigure{
        \includegraphics[width=60mm,clip]{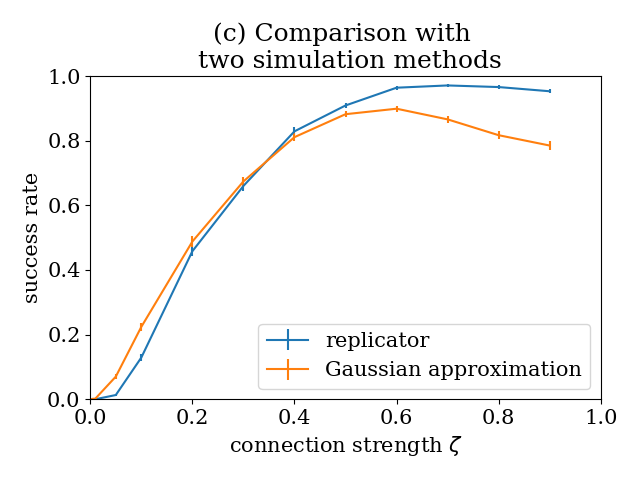}
        }
        \subfigure{
        \includegraphics[width=60mm,clip]{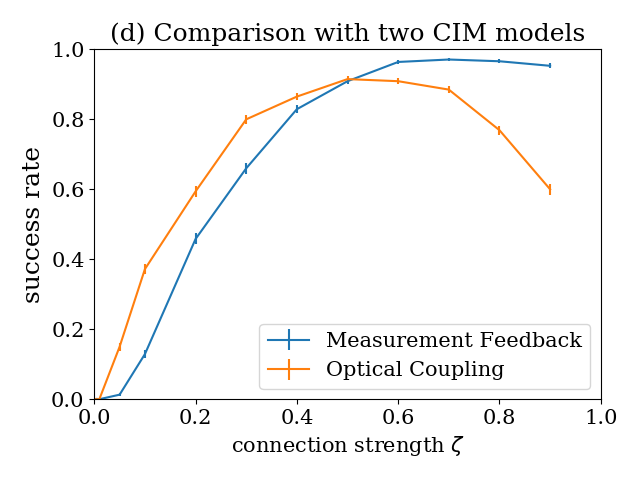}
        }
        \caption{(Color online) (a) Success rate vs. connection strength $\zeta$ for the exact replicator dynamics model (blue (dark gray) online)  and Gaussian approximation (orange (light gray) online). (b) Success rate vs. connection strength $\zeta$ for the measurement-feedback-based CIM (blue (dark gray) online) and optical delay line coupling CIM. (orange (light gray) online). (c) Success rate comparison; the same as (a) with two simulation methods in the case of $\kappa = 0.01$ (d) Success rate comparison with two CIM models in the same condition with (c).} \label{fig:linear1000}
\end{figure}
 
  To reveal the unique capability of the CIM as an optimizer, we simulated N=16 DOPOs coupled by the nearest neighbor antiferromagnetic interaction in a one-dimensional ring configuration:
 \begin{eqnarray}
 J_{ij} = \left\{ \begin{array}{llr} -1, &&|i-j| = 1, \\ 0, && otherwise. \end{array} \right.
 \end{eqnarray}
The two degenerate ground states of this model are $(-1,1,-1,...,-1,1)$ and $(1,-1,1,...,1,-1)$.

 We compared three models. The first and second models are based on the exact replicator dynamics and the Gaussian approximation. The third model is based on the quantum theory of an optical delay line coupling CIM analyzed by \cite{TakataPRA}\cite{Maruo}.

  Figure 4(a) shows the success rates of finding either one of the two degenerate ground states in 1000 trials, where $\gamma_s = 1, \gamma_p = 10, \kappa = 0.1, \xi = 0.1$ and the injection rate $\zeta$ changes from $0.01$ to $1$. $\epsilon_{p}$ is linearly increased from 0 to 1.2 times $\epsilon_{th}$. 
  
When the mutual coupling parameter $\zeta$ is small, the potential landscape for the DOPO field is almost symmetric with respect to $X=0$ as shown in Fig. 6. In such a case, the measurement-induced wavepacket reduction and the feedback-induced wavepacket displacement play major roles in the solution search process. In this case, the tightly confined Gaussian wavepacket is more advantageous than the broadly spread exact wavepacket, because the latter introduces more random measurement results and takes a longer time to reach a final decision. However, when the mutual coupling parameter $\zeta$ is large, the potential landscape of the DOPO field is highly asymmetric. In such a case, the decision making process in the Gaussian approximation happens too quickly so that the system is easily trapped in a wrong state as shown in Fig.5(b). The non-Gaussian wavepacket induced quantum tunneling in the exact replicator dynamics plays an important role for the system to escape from the wrong solution. In this case, the broadly spread exact wavepacket is more advantageous than the tightly confined Gaussian wavepacket, as shown in Fig. 5(b). The numerical simulation results in Fig.4(a) confirm this trade-off relation and also suggest the importance of quantum tunneling\cite{Kimsler} in the solution search process of the CIM near the threshold.

\begin{figure}[htbp]

        \subfigure{
        \includegraphics[width=60mm,clip]{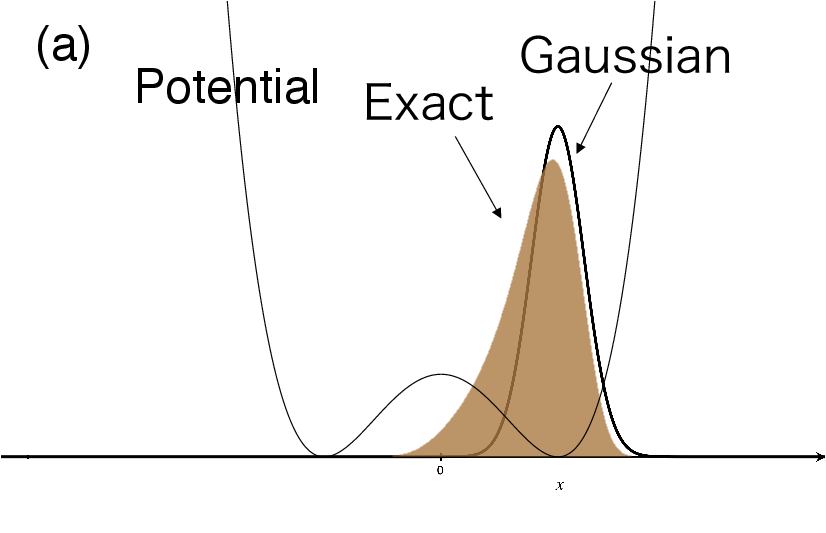}
        }
        \subfigure{
        \includegraphics[width=60mm,clip]{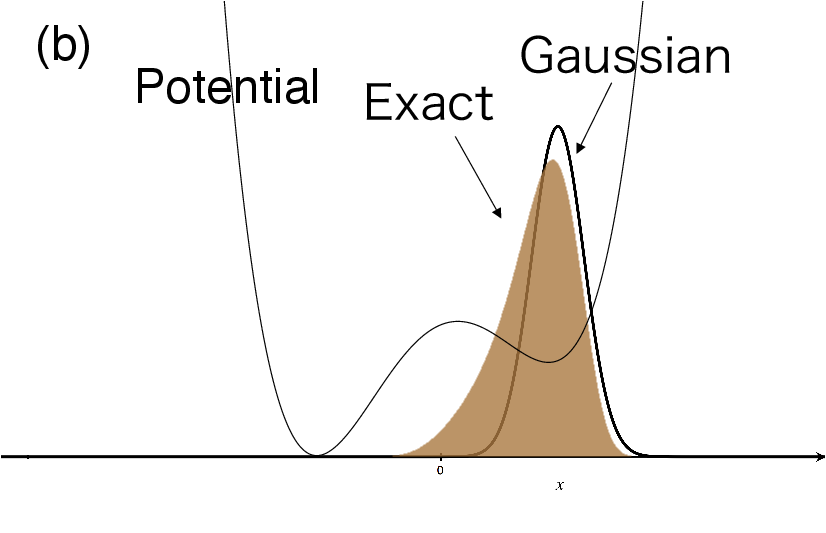}
        }
        \caption{(Color online) (a) Almost symmetric potential for the DOPO field and two wavepackets with a small $\zeta$ value. (b) Asymmetric potential for the DOPO field with a large $\zeta$ value.}
\end{figure}

The numerical results in Fig.4(b) shows that the optical coupling CIM is more efficient than the measurement-feedback-based CIM when $\zeta$ is small. However, in the case of $\zeta > 0.6$, the measurement-feedback-based CIM has a higher success rate than the optical delay line coupling CIM. When the connection strength $\zeta$ is close to one, it means that the extracted signal field is boosted by a high-gain phase sensitive amplifier before it is injected back to the DOPO cavity (Fig.1 of ref.\cite{Maruo}). This is necessary since the injection coupler has a very small coupling constant. During this external amplification process, the vacuum fluctuation added to the extracted signal field is also amplified and contributes to the degradation of the degree of negative correlation among neighboring DOPOs. Because of this reason, there is an optimum coupling strength to maximize the degree of correlation in the optical delay line coupling CIM (see Fig.3(b) of ref.\cite{TakataPRA}). The maximum success rate at $\zeta \simeq 0.5$ corresponds to this optimum coupling strength. In the case of the measurement-feedback-based CIM, the search mechanism is not the formation of correlation between DOPOs but the feedback signal-induced quantum tunneling so that a higher coupling strength $\zeta$ always improves the success rate.

Fig.4 (c) and (d) shows the same comparison as Fig.4 (a) and (b) where $\kappa = 0.01$ and $g=0.002$, which means that non-linear coupling constant $\kappa$ and the saturation parameter g are ten times smaller. The tendency of the results are not so different from the case of $\kappa = 0.1$ and $g=0.02$. The success rate of $\kappa = 0.01$ and $g=0.002$ case is higher than that of $\kappa = 0.1$ and $g=0.02$ case,  Therefore, we can conclude that the saturation parameter $g$ plays a crucial role in the computation.

\section{Conclusion}
We presented the two theoretical methods for numerically simulating the measurement-feedback-based CIM with exact replicator dynamics and Gaussian approximation.

The Gaussian approximation method is computationally inexpensive, that is, we can efficiently implement this model in modern digital computers, while the exact replicator dynamics is computationally expensive because we have to run many Brownian particles simultaneously to properly account for the non-unitary state reduction of non-Gaussian wavepackets. As we have shown, the exact replicator dynamics predicts a better performance than the Gaussian approximation when the mutual coupling strength is large. Therefore, the present result supports the current experimental effort to develop an actual physical device solving replicator dynamics\cite{Science1}\cite{Science2} rather than using the CSDE under Gaussian approximation as a new algorithm.

\section*{Acknowledgement}
This research was funded by ImPACT Program of Council for Science, Technology and Innovation (Cabinet Office, Government of Japan). The authors wish to thank P. Drummond, H. Mabuchi, R. Hamerly and A. Yamamura for their critical discussions.

\end{document}